\documentclass[letterpaper]{article}
\usepackage{aaai}
\usepackage{times}
\usepackage{helvet}
\usepackage{courier}
\usepackage{amsmath,graphicx,amssymb,subfig, booktabs}
\usepackage{lettrine}
\usepackage{textcomp}
\begin{document}
%
\title{Content-based Video Indexing and Retrieval Using Corr-LDA}
\author{\bf Rahul Radhakrishnan Iyer$^1$,  {\bf Sanjeel Parekh$^2$}, {\bf Vikas Mohandoss$^3$}, {\bf Anush Ramsurat$^4$}, {\bf Bhiksha Raj$^1$} and {\bf Rita Singh$^1$} \\ \\
 $^1$ Language Technologies Institute, School of Computer Science, Carnegie Mellon University, Pittsburgh, PA, USA \\ $^2$ Technicolor, Cesson-S\'evign\'e, France \\ $^3$ Information Technology, NITK Surathkal, Mangalore, KA, India \\ $^4$ Computer Science and Engineering, NIT Tiruchirappalli, Tiruchirappalli, TN, India\\
}

\maketitle
\begin{abstract}
\begin{quote}
Existing video indexing and retrieval methods on popular web-based multimedia sharing websites are based on user-provided sparse tagging. This paper proposes a very specific way of searching for video clips, based on the content of the video. We present our work on Content-based Video Indexing and Retrieval using the Correspondence-Latent Dirichlet Allocation (corr-LDA) probabilistic framework. This is a model that provides for auto-annotation of videos in a database with textual descriptors, and brings the added benefit of utilizing the semantic relations between the content of the video and text. We use the concept-level matching provided by corr-LDA to build correspondences between text and multimedia, with the objective of retrieving content with increased accuracy. In our experiments, we employ only the audio components of the individual recordings and compare our results with an SVM-based approach. 
\end{quote}
\end{abstract}


\noindent \section{Introduction}
\label{sec:intro}

Multimedia data typically entails digital images, audio, video, animation and graphics together with text data. Multimedia content on the internet is growing at an exponential rate. It is for this reason that indexing and retrieval of multimedia has become a pertinent issue and a hot topic of current research. Video indexing and retrieval have a wide spectrum of promising applications, motivating the interest of researchers worldwide. Our work proposes a novel solution to the problem of video indexing and retrieval. It is motivated by the premise that current multimedia indexing and retrieval techniques, which are largely based on sparse-tagging, often lead to erroneous and completely unwarranted results in terms of accuracy of video clippings returned to the user based on the search query. In order to provide better results, we base our video search paradigm on the actual content of the recordings.   


An annotated video has a dual representation comprising both sensory words, which are individualized components of the multimedia file under consideration, as well as textual words. In mathematical terms, given a search query, the goal of video retrieval is to find the probability of retrieving a video given a search query. The objective of the video indexing problem is, given an unannotated video, we require the probability of a particular text descriptor being associated with that video. This model allows for automatic textual annotation of multimedia for indexing and retrieval. We assign a score based on probability of each word in the query matching with a video in our database, and rank videos accordingly, retrieving the ones with scores above a set threshold. 
We use a measure called ‘perplexity’, as first developed by Blei et al [2], to quantize the annotation quality of a video.

\subsection{Related Work}
Video query by semantic keywords is one of the most difficult problems in multimedia data retrieval. The difficulty lies in the mapping between low-level video representation and high-level semantics. In \cite{naphide2001probabilistic}, the multimedia content-access problem is formulated as a multimedia pattern recognition problem. A probabilistic framework is proposed to map the low-level video representation to high level semantics using the concepts of \textit{multijects} and their networks called \textit{multinets}. 

A widely used structure for a stream of video data is to store them as contiguous groups of frames called segments. Each segment comprises of collections of frames. Usually, video comparisons compare each frame to find the similarity between two video streams. Research conducted in this domain has mostly dealt with structural analysis of video, which includes methods such as shot boundary detection \cite{zhang2001automatic}, key frame extraction and scene segmentation, extraction of features including static key frame features, object and motion features, video annotation and retrieval.  A concise review of content-based retrieval and indexing techniques is provided in \cite{geetha2008survey,hu2011survey}.


In recent years research has focused on the use of internal features of images and videos computed in an automated or semi-automated way \cite{Fablet00statisticalmotion-based}. Automated analysis calculates statistics, which can be approximately correlated to the content features. This is useful as it provides information without costly human interaction.

The common strategy for automatic indexing had been based on using syntactic features alone. However, due to its complexity of operation, there has been a paradigm shift in the research concerned with identifying semantic features \cite{fan2004classview}. User-friendly Content-Based Retrieval (CBR) systems operating at semantic level would identify motion-features as the key besides other features like color, objects etc., because motion (either of camera motion or shot editing) adds to the meaning of the content. The focus of present motion based systems had been mainly in identifying the principal object and performing retrieval-based on cues derived from such motion. With the objective of deriving semantic-level indices, it becomes important to deal with the learning tools. The learning phases followed by the classification phase are two common envisioned steps in CBR systems. Rather than the user mapping the features with semantic categories, the task could be shifted to the system to perform learning (or training) with pre-classified samples and determine the patterns in an effective manner. 
\newline{}

In the past several researchers have considered the problem of building semantic relations or correspondences for modeling annotated data. We mention here several important contributions to this area. In \cite{Barnard:2003:MWP:944919.944965} authors investigate the problem of auto-annotation and region naming for images using Mixture of Multi-Modal Latent Dirichlet Allocation (MoM-LDA) and Multi-Modal Hierarchical Aspect Models. 

Use of Canonical correlation analysis (CCA) for joint modeling is proposed in \cite{Rasiwasia:2010} and \cite{Hodosh:2013}. The idea is to map both images and text into a common space and then define various similarity metrics. This can be used for both indexing and retrieval. Kernel CCA is used by \cite{Hodosh:2013} where the problem is formulated as that of maximizing an affinity function for image-sentence pairs. Rank SVM based approach is proposed in \cite{6595890} where linear classifiers are trained for an image with relevant and irrelavant captions. 

Many other novel methods have been proposed for establishing relations between images and text \cite{Carneiro,Farhadi,Gupta,Datta}. The Labeled LDA model is introduced by \cite{Ramage:2009:LLS:1699510.1699543} to solve the problem of credit attribution by learning word-tag correspondences. Recently, several approaches involving machine learning, neural networks and deep learning have also been used in the visual and language domains \cite{vinyals2014show,fang2014captions,karpathy2014deep,donahue2014long,mao2014explain,li2016joint,iyer2018transparency,li2018object,gupta2016analysis,honke2018photorealistic,iyer2017detecting}.\\

\noindent  We propose to use an extension of LDA \cite{BleiLDA} called the Correspondence-LDA, familiarized by Blei and Jordan \cite{BleiCorr} to bridge the gap between low level video representation and high level semantics comprehensively. 
Our approach is significantly different from the discussed approaches because we model the problem in the probabilistic framework where both captions and videos are said to be generated from a set of topics. Moreover, we use a bag-of-words representation for both video content and text. Particularly, we differ from Blei's usage of Corr-LDA for image annotation and retrieval in the following two aspects:

\begin{enumerate}
\item  Blei et al. segment an image into different regions and the feature vectors for each of those regions are computed. We do not perform any such segmentation for videos

\item We use the bag-of-words representation for feature vectors. As a consequence, we use Multinomial distribution where as Blei et al.  assume that visual words are sampled from a Multivariate-Gaussian distribution

\end{enumerate}
\subsection{Paper Organization} 
The paper is organized as follows. We formulate the problem and explain the technical approach in sections (\ref{secprobform}) and (\ref{sec:techappr}) respectively. The experimental results obtained for a set of videos and their implications are discussed in section (\ref{sec:results}). We present a few limitations of our work in section (\ref{sec:limits}) and finally draw conclusions and explore possibilities of future work in the last section (\ref{sec:concl}) of the paper.

\section{Problem Formulation}
\label{secprobform}
We assume a collection of videos, $\Omega$, a fixed-size textual dictionary $D^{Dict} = [D_1, D_2, \ldots ,D_n]$ and a sensory word dictionary $S^{Dict} = [s_1, s_2,\ldots ,s_m]$ . The textual dictionary is built from a list of all possible words for search queries, indexed alphabetically. An annotated video $V_i \in \Omega$, then, has a dual representation in terms of both the sensory words $F^{i} = [f_1, f_2,\ldots ,f_m]$, which are individualized components of the multimedia file under consideration, and textual descriptors $D^{i} = [D_{i_1}, D_{i_2}, \ldots ,D_{i_k}]$, where $F^{i}$ represents the frequency of occurrence of the sensory words in the video $V_i$ i.e. $f_j$ denotes the frequency of occurrence of $s_j$ in the video $V_i$ considered, and $D^{i}$ represents the annotation of the video $V_i$, where $D_{i_j} \in D^{Dict}$, for $j = 1, 2, \ldots, k$. The relevant parameters required for the tasks mentioned below are estimated from the model. The tasks can be clearly defined as : 
\begin{itemize}
\item{\textit{Video Retrieval}: This refers to the task of retrieving videos relevant to a text search query. In formal terms, for a query $q$, we determine $P(q|V)$ $\forall$ $V$ $\in \Omega$; this represents the probability of the query being associated with a given video. We rank these set of videos based on the obtained set of probabilities and retrieve the ones above a set threshold.} 

\item{\textit{Video Indexing}: This task refers to annotating a video that is without captions. Formally, given an unannotated video $V^*$ $\not\in \Omega$ , we estimate $P(D_i|V^*)$ $\forall$ $D_i$ $\in D^{Dict}$, after which we rank the textual descriptors in decreasing order of probabilities and return the ones above a threshold, which best describe the video in consideration.}
\end{itemize}

\section{Technical Approach} 
\label{sec:techappr}
In this paper, we approach the problem in the \textit{Correspondence Latent Dirichlet Allocation} (Corr-LDA) framework. The main idea is to learn correspondences between multimedia (video in this case) and text. We later use the learned statistical correspondence to retrieve videos in response to text queries, by computing : 
\begin{itemize}
\item $P(V,D)$ : Required for clustering and organizing a large database of videos.
\item $P(D|V)$ : Required for automatic video annotation and text-based video retrieval.
\end{itemize}
The tool which facilitates in creating these correspondences is the Corr-LDA. In order to explain our approach first we introduce certain necessary terminology in the next section.


\subsection{Preliminaries} 
A \textit{corpus} is a collection of documents. A \textit{document} is defined as a collection of words. The generative probabilistic framework used to model such a corpus is the Latent Dirichlet Allocation (LDA). It models documents as mixtures over latent topics, where each topic is represented by a distribution over words (a histogram).

Video files and textual descriptors are represented as \textit{bags of words} (BoW) in our work. Just as a textual document consists of words, a multimedia document consists of \textit{sensory words}. This model is a simplifying assumption used in natural language processing and information retrieval wherein, a text (such as a sentence or a document) is represented as an unordered collection of words, disregarding grammar and even word order. From a data modeling point of view, the bag-of-words model can be represented by a co-occurrence matrix of documents and words as illustrated. \\

\begin{figure}
\centering
\includegraphics[scale=0.4]{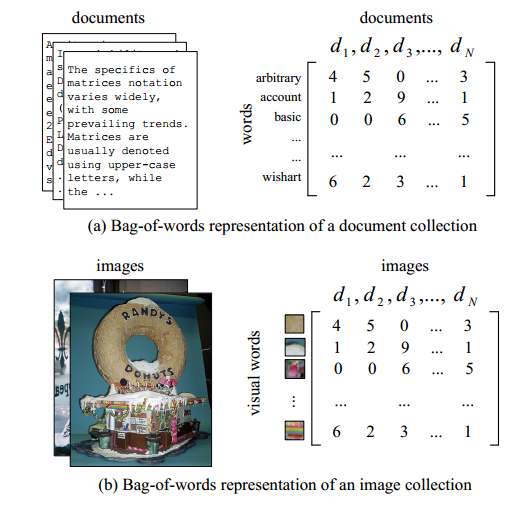}
\caption{Understanding \textit{Bag-of-words} representation for multimedia \cite{Xiao2011}}
\label{bow}
\end{figure}

\noindent Thus, both the multimedia in consideration and its textual description have BoW representation. We also represent the textual search query as a BoW. 

The Corr-LDA is a generative probabilistic model that is an extension of LDA for modeling documents that consist of pairs of data streams. In particular, one data type can be viewed as an annotation of the other data type. Examples of such data sets include images and their associated captions, papers and their bibliographies etc. Like LDA, Corr-LDA can be viewed in terms of a probabilistic generative process that first generates the region descriptions and subsequently generates the annotation words. Here, we modify the original Corr-LDA model \cite{BleiCorr} by replacing the Gaussian distribution by multinomial distribution, which leads to a new generative process and parameter estimation algorithm. In particular, for images, it first generates region descriptions from an LDA model (with multinomial distribution in this case). Then, for each of the annotation words, one of the regions is selected from the video/image and a corresponding annotation word is drawn, conditioned on the factor that generated the selected region. 

Under the Corr-LDA model, the regions of the image/video can be conditional on any ensemble of factors but the words of the caption must be conditional on factors, which are present in the video/image. Also, the correspondence implemented by Corr-LDA is not a one-to-one correspondence, but is more flexible: all caption words could come from a subset of the video/image regions, and multiple caption words can come from the same region.

Corr-LDA can provide an excellent fit of the joint data as well as an effective conditional model of the caption given a video/image. Thus, it is widely used in automatic video/image annotation, automatic region annotation, and text-based video/image retrieval.  \\

\begin{figure}
 \centering
 \includegraphics[scale=0.2]{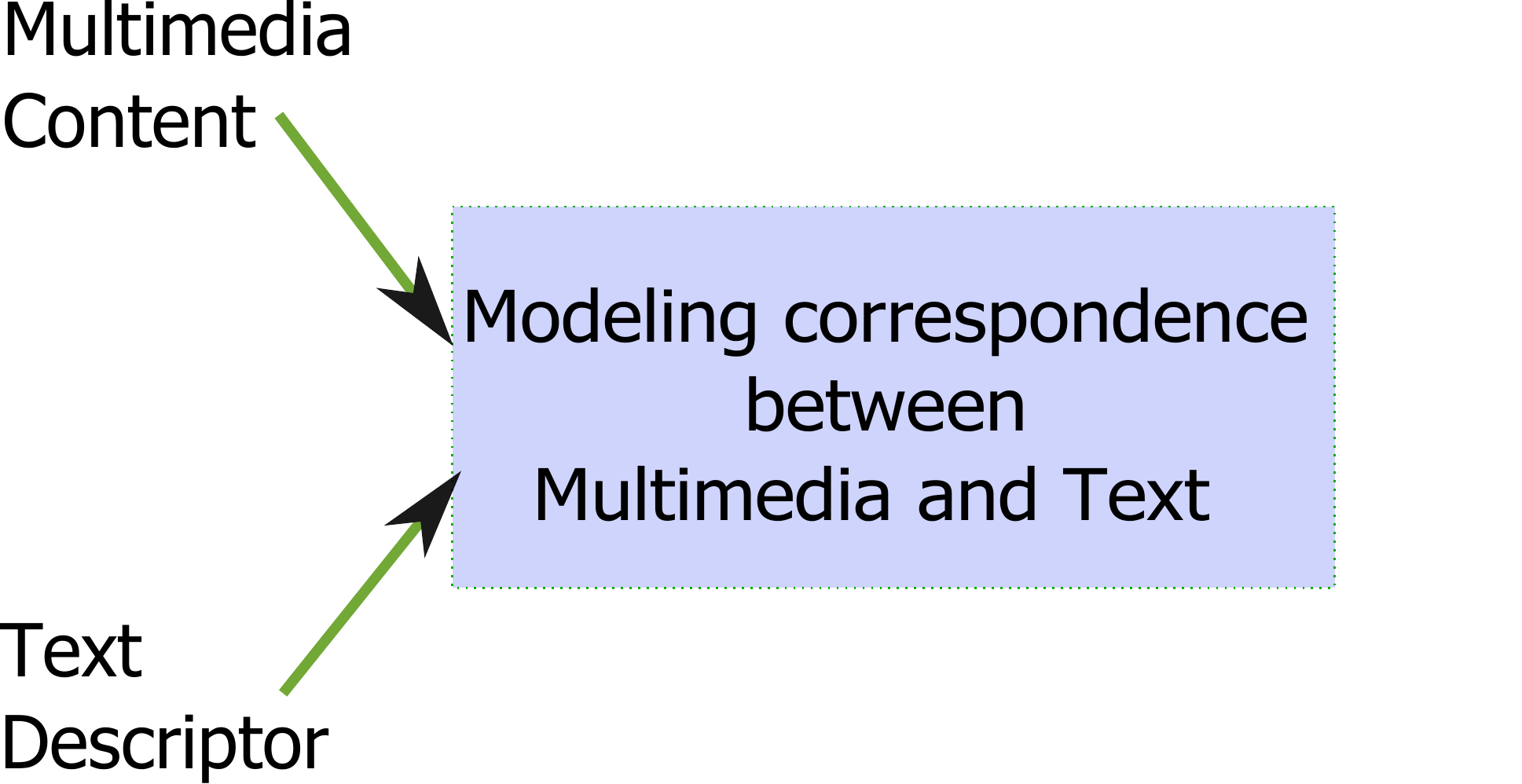}
 \caption{The crux of Corr-LDA: Corr-LDA is used to model the correspondence between audio/video and text.}
 \label{crux}
\end{figure}

\begin{figure}
\centering
\includegraphics[scale=0.4]{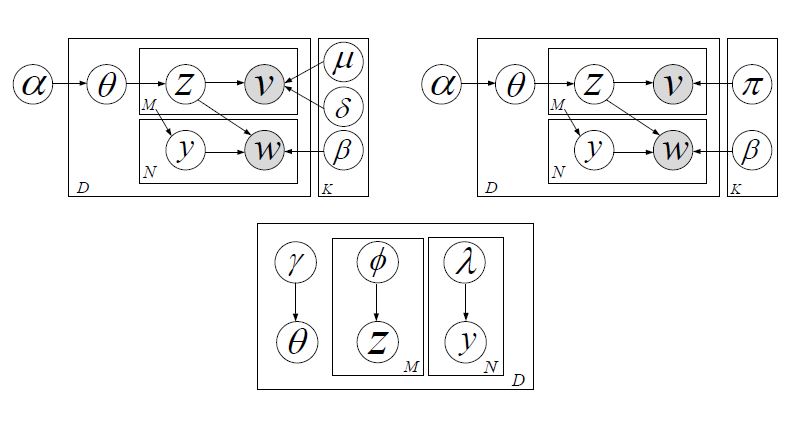}
\caption{Graphical model representations: (\textbf{Left}) Original Corr-LDA \cite{BleiCorr}. Nodes represent random variables; shaded nodes are observed random variables, unshaded nodes are latent random variables. Edges denote possible dependence between random variables; plates denote replicated structure. Note that the variables $y_n$ are conditioned on M. (\textbf{Right})  Modified Corr-LDA. (\textbf{Bottom})  Variational distribution used to approximate the posterior in Corr-LDA.  \cite{Xiao2011}}
\label{CorrLDA-modified}
\end{figure}

\noindent The problem at hand is learning correspondence information between videos and text. In other words, we need to estimate the probability of $p(w|V∗)$, $p(V|w)$, where \textit{V*} represents the concerned video and \textit{w} represents text. Using the modified version of the Corr-LDA model as described below, we apply it to model audio/text and videos/text. We fix the number of topics \textit{T} in the procedure. Formally, the generative procedure can be expressed through the following sequence pictorially depicted in Fig. \ref{CorrLDA-modified}\footnote{We have acquired permission from Han Xiao for reproducing Fig. \ref{bow} and Fig. \ref{CorrLDA-modified} from his master's thesis \cite{Xiao2011}.}:

\begin{enumerate}
\item {Sample topic distribution $\theta \sim Dirichlet(\alpha)$}
\item{For each sensory word $v_m: m \in (1, \ldots,  M)$

\begin{itemize}
\item {Draw topic assignment $z_m|\theta \sim Multinomial(\theta)$}
\item {Draw sensory word $v_m|z_m \sim Multinomial(\Pi_{z_m})$}
\end{itemize}}

\item{For each textual word $w_n; n \in (1 \ldots N)$
\begin{itemize}
 \item {Draw discrete indexing variable $y_n \sim  Uni(1,\ldots, M)$}
\item {Draw textual word $w_n \sim Multinomial(\beta_{z_{y_{n}}})$}

\end{itemize}}

\end{enumerate}
\begin{figure}
 \centering
 \includegraphics[scale=0.5]{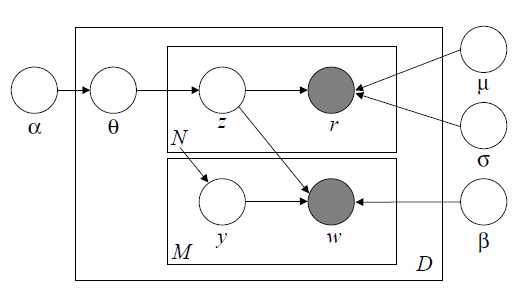}
 \caption{Graphical Representation for Corr-LDA \cite{BleiCorr}}
\label{CorrLDA}
\end{figure}

Note that in Fig. \ref{CorrLDA-modified} (\textbf{Bottom}), the variational graphical model creates independency between $z$, $\theta$ and $y$, which enables us to factorize the joint probability.

\subsection{Approach}
\begin{figure}
 \centering
 \includegraphics[scale=0.4]{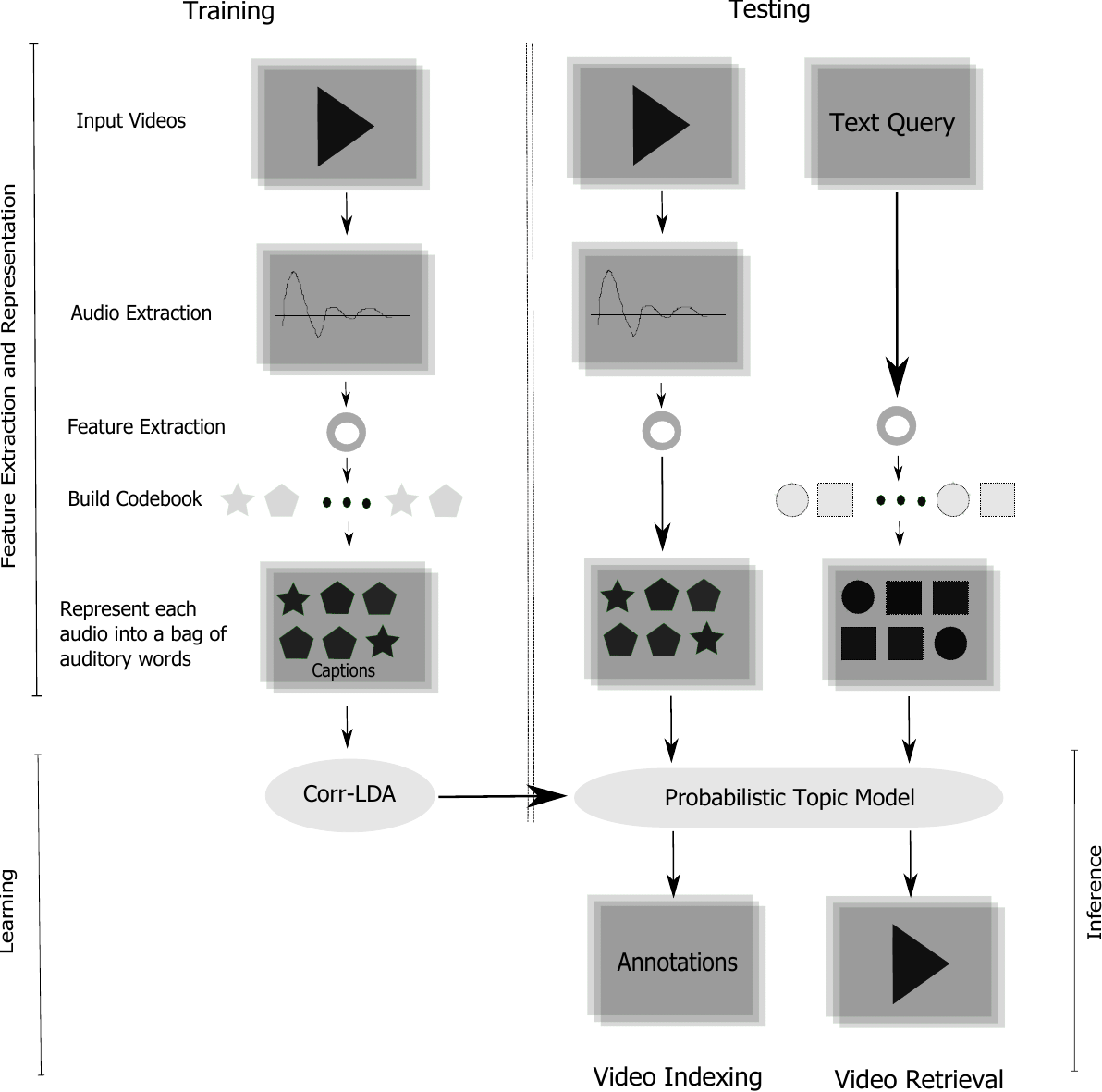}
 \caption{Algorithm: Flowchart Representation}
 \label{algo-flow}
 \end{figure}

Here, our corpus is a collection of videos, a document is a video and the annotations refer to the descriptors. Words can refer to both \textit{auditory words} as well as textual words (BoW). Auditory words are the sensory words obtained from the audio components of the recordings. A concise representation of the proposed algorithm is presented in Fig. \ref{algo-flow}.

The model is trained on a training set of videos. Using the parameters estimated from the trained model, we perform the required actions on test inputs to output required results. Exact inference of the parameters of this model is intractable. So, we use an approximate inference algorithm called \textit{Variation Expectation Maximization (EM)} to estimate the parameters. Due to space constraints, explanation of the inference algorithm has not been included. The reader is referred to \cite{Xiao2011} for details \footnote{A technical note can also be referred to at http://home.in.tum.de/\texttildelow{}xiaoh/pub/derivation.pdf }. We use these estimated parameters from the learning phase for indexing and retrieval of videos. 


The parameters estimated from the learning phase to accomplish the tasks of video indexing, retrieval and calculation of perplexity are (refer to Fig. \ref{CorrLDA-modified}) :
\begin{enumerate}
\item $\theta$ : Represents the topic-document distribution
\item $\beta$ : Represents the topic-word distribution
\end{enumerate}


\begin{itemize}

\item{\textit{Video Retrieval}: 

Given a query consisting of $n$ words  $q = [q_1, q_2, \ldots ,q_n]$, we compute a score for each video $V_i \in \Omega$ and rank them accordingly. First we remove all the \textit{stop-words} (taken from NTLK) from the search query. Then the scores can be computed as follows:
\begin{equation}
Score_{i}=\prod\limits_{n} P(q_n|V_i)
=\prod\limits_{n} \sum\limits_{z} P(z|V_i) P(q_n|z)
\label{vidretr1}
\end{equation}}


Here $z$ is the latent variable in the model, which corresponds to the notion of a \textit{topic}. We pick a video $V_i$ from our database $\Omega$ and compute the probability of each word of the search query being associated with that video. The total probability of the search query itself being associated with $V_i$ is the product of individual probabilities calculated in (\ref{vidretr1}). The probability of each individual word being associated with that video $V_i$ can be calculated as follows. If we assume that topic \textit{z} has been chosen, then the probability that the word $q_j$ is associated with that topic is $P(q_j|z)$ (given by $\beta$). This multiplied with the probability of picking topic \textit{z} when considering video $V_i$ (given by $\theta$) gives the probability that $q_j$ comes from topic \textit{z} when video $V_i$ is considered. But the word $q_j$ could have come from any topic. So, considering all possibilities, we take the sum of above probabilities $\forall$ $z$. This is done for all words $q_j$ in the search query $q$ and their products are taken to compute the score for the $i^{th}$ video $V_i$.  
This gives the score for $V_i$. We calculate such scores $\forall$ $V_i$ $\in \Omega$ and rank the videos based on that. 




\item{\textit{Video Indexing}: 

Using the notation from the previous section, $P(D_j|V^*)$ provides us with relevant annotations for the video $V^*$ (need not $\in \Omega$), $D_j \in D^{Dict}$, where $D^{Dict}$ refers to the textual dictionary. We compute scores for each word $D_j$ in the textual dictionary as follows :
\begin{equation}
Score_j = P(D_j|V^*) = \sum\limits_{z} P(z|V^*) P(D_j|z) 
\end{equation}

We calculate the score assuming that the word $D_j$ is taken from the topic \textit{z} and do it $\forall$ $z$. Thus, \textit{$Score_j$} gives the score for the $j^{th}$ word in the textual dictionary being associated with $V^*$. This is done $\forall$ word $D^j$ in the dictionary. Next, we rank the words in decreasing order of score and return the ones, which have scores above a set threshold. These words are the obtained annotation for the untagged video.


}
\end{itemize}

\section{RESULTS}
\label{sec:results}
\subsection{Experimental Results}

Our experiments were conducted on the MED11 example-based multimedia-event retrieval task \cite{dataset}. The task here is as follows: we are given a collection of multimedia recordings. We are also provided a small number of examples of recordings representing a category of events such as \textit{repairing an appliance}, \textit{parade}, \textit{parkour}, or \textit{grooming an animal}. The objective is to retrieve all other instances of the same category from the data set, based on what can be learned from the example recordings. In our experiment we only employed the audio components of the individual recordings to perform the retrieval; the visual features were not used. One of the reasons for this is to explore how well an audio can describe a recording. Furthermore, considering only the audio features instead of visual ones in our experiments speeds up execution, saving space and time, without compromising much on the quality. The method used for generating audio bag-of-words is as described in \cite{chaudhuri2011}. Herein, an audio is represented as a set of smaller units called ``Acoustic Unit descriptor" (AUD) and the BoW representation is computed over sequences of AUDs. We have used Han Xiao's Corr-LDA implementation for the experiments.

To compare our results for retrieval, we use SVM-based classifiers. We train two sets of SVM classifiers, each with five classifiers: one for each of the video categories, using 500 training videos from our dataset. In both the cases, we train the SVMs on visual features, considering 10 equally spaced frames per video. In one case, referred to as SVM-CTS, we extract 26 low level visual features pertaining to color, texture and structure for each frame \cite{sergyan2008color}. In the second case, we extract dense SIFT features, 1536 per frame, with a step size of 250. This model is referred to as SVM-DSIFT. 

The models, both Corr-LDA and SVMs, are trained on a set of 500 videos and tested on a set of 125 mixed category videos.  



\begin{table*}
\vspace{2ex}
\begin{tabular}{l | c c | c c |c c| c c |c c}
\toprule
\textbf{Category / Query} & \multicolumn{2}{c}{\textbf{SVM-DSIFT}} & \multicolumn{2}{c}{\textbf{SVM-CTS}} & \multicolumn{2}{c}{\textbf{Corr-LDA-5}} & \multicolumn{2}{c}{\textbf{Corr-LDA-50}} & \multicolumn{2}{c}{\textbf{Corr-LDA-100}}\\

& Precision & Recall & Precision & Recall & Precision & Recall & Precision & Recall & Precision & Recall\\


\midrule
Grooming Animal & 0.600 & 0.220 & 0.500 &0.185 &0.300 & 0.111& 0.600 & 0.222 &0.700 &0.259\\
Making Sandwich & 0.300  & 0.125 & 0.300 & 0.125 & 0.600 &0.250 &0.600 &0.250 &0.300 &0.125\\
Parade & 0.200  & 0.080 & 0.600 & 0.240 & 0.300 &0.120 &0.800 &0.320 &1.000 &0.400\\
Parkour & 0.100  & 0.042 & 0.800 & 0.333 &0.500 &0.208 &0.600 &0.250 &0.700 &0.292\\
Repair Appliance & 0.600  & 0.240 & 0.600 &0.240 &1.000 &0.400 &1.000 &0.400 &1.000 &0.400\\

\bottomrule
\end{tabular}
\centering
\caption{Comparison of precision and recall at 10 for SVM Based Classifiers and Corr-LDA (for different topics) for Retrieval}
\label{pr10}
\end{table*}


\begin{table}
\vspace{2ex}
\begin{tabular}{l | l | l | l}
\toprule
& \multicolumn{3}{c}{\textbf{Topics}}\\
\textbf{Parameter} & 5 & 50 & 100\\
\midrule
Number of Iterations &  67 & 639 & 621\\
Convergence Threshold & $10^{-7}$ & $10^{-7}$ & $10^{-7}$\\
Dirichlet Prior $\alpha$ of Corr-LDA & 0.1 & 0.2 & 0.2\\
Number of Topics & 5 & 50 & 100\\
Number of Training Videos & 500 & 500 & 500\\
Number of Test Videos & 125 & 125 & 125\\
Number of Auditory Words & 8193 & 8193 & 8193\\
Number of Textual Words & 9654 & 9654 & 9654\\
\bottomrule
\end{tabular}
\centering
\caption{Experimental Parameters}
\end{table}

\noindent The following results were observed:
\begin{itemize}
\item \textit{Video Retrieval} :
\begin{itemize}
\item We compute precision @ 10, recall @ 10 and mean average precision @ 10 (MAP) for both the models: Corr-LDA trained with 5, 50 and 100 topics (referred to as Corr-LDA-5, Corr-LDA-50, Corr-LDA-100), and SVM-based classifiers. MAP is computed over the queries mentioned in Table \ref{pr10}.
\item As can be seen from Tables \ref{pr10} and \ref{MAP}, considering each SVM category as an input query,  Corr-LDA clearly outperforms the SVM based frame-by-frame classification. We observe from Table \ref{MAP} that the performance of Corr-LDA depends on the number of topics. SVM-CTS performs better when Corr-LDA is trained with just five topics. 
\item It is worth noting that although the SVMs used the features from the video itself, Corr-LDA having used just the audio components of the recordings, still outperformed.
\item It is also observed that as the number of topics are increased, the accuracy of retrieval of videos based on the search query increases. This is particularly evident with multi-word queries.
\item From Figure \ref{prc} we see that for two categories namely \textit{parade} and \textit{repair appliance} all of the first few results retrieved using Corr-LDA (100 topics) are relevant. However, the precision is poor for categories \textit{grooming animal} and \textit{making sandwich}. In this respect, retrieval is poorer for SVM-based systems. 
\item This shows that having a higher number of topics helps the Corr-LDA system to learn, differentiate, segregate videos better and hence create concept-level matching with increased accuracy.
\end{itemize}

\begin{table}
\vspace{2ex}
\begin{tabular}{l | c}
\toprule
\textbf{Model} & \textbf{Mean Average Precision}\\


\midrule
SVM-DSIFT & 0.121\\
SVM-CTS & 0.160 \\
Corr-LDA-5 & 0.151  \\
Corr-LDA-50 &0.231 \\
Corr-LDA-100 & 0.243 \\

\bottomrule
\end{tabular}
\centering
\caption{Comparison of MAP at 10 for different models}
\label{MAP}
\end{table}


\begin{figure*}[!t]
\centering
\subfloat[Corr-LDA with 100 topics]{\includegraphics[width=2.2in]{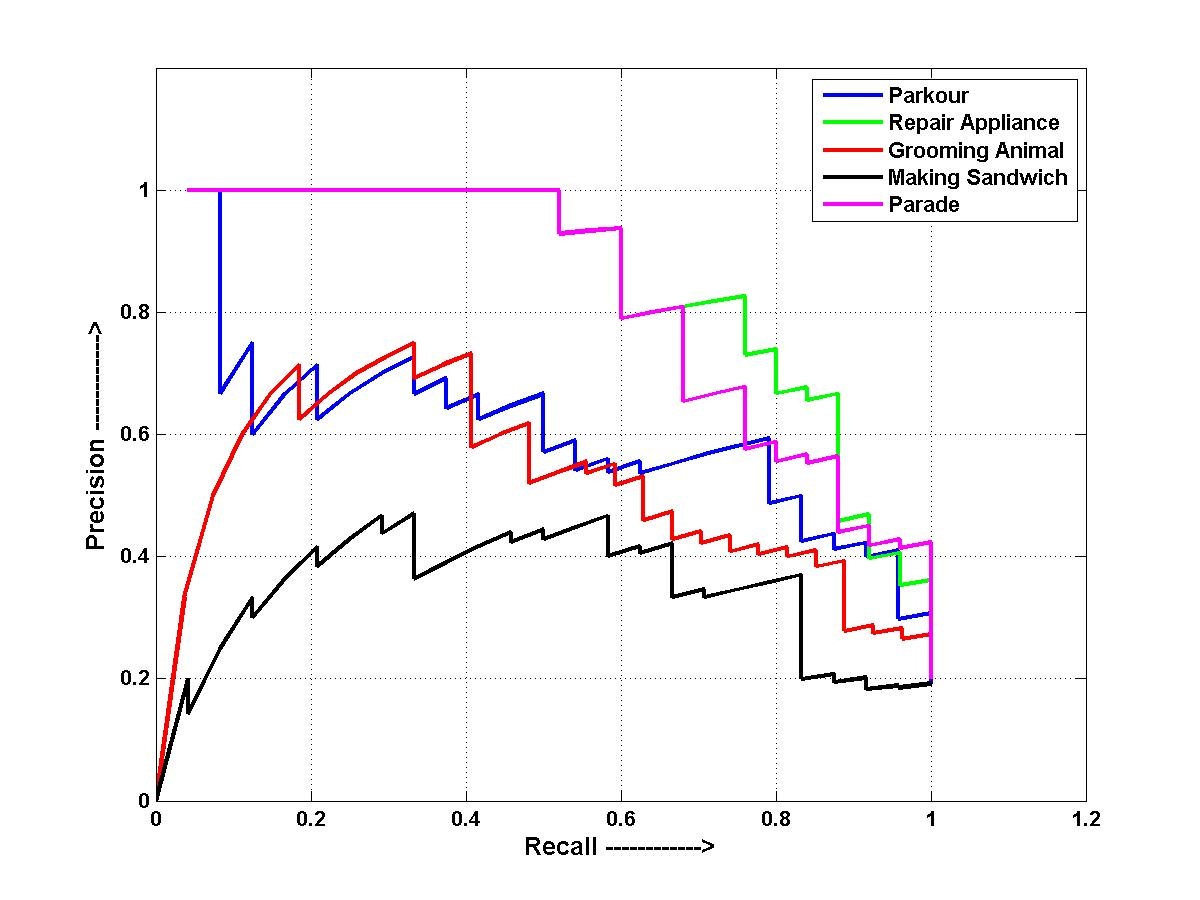}%
\label{lda100}}
\hfil
\subfloat[SVM-DSIFT]{\includegraphics[width=2.2in]{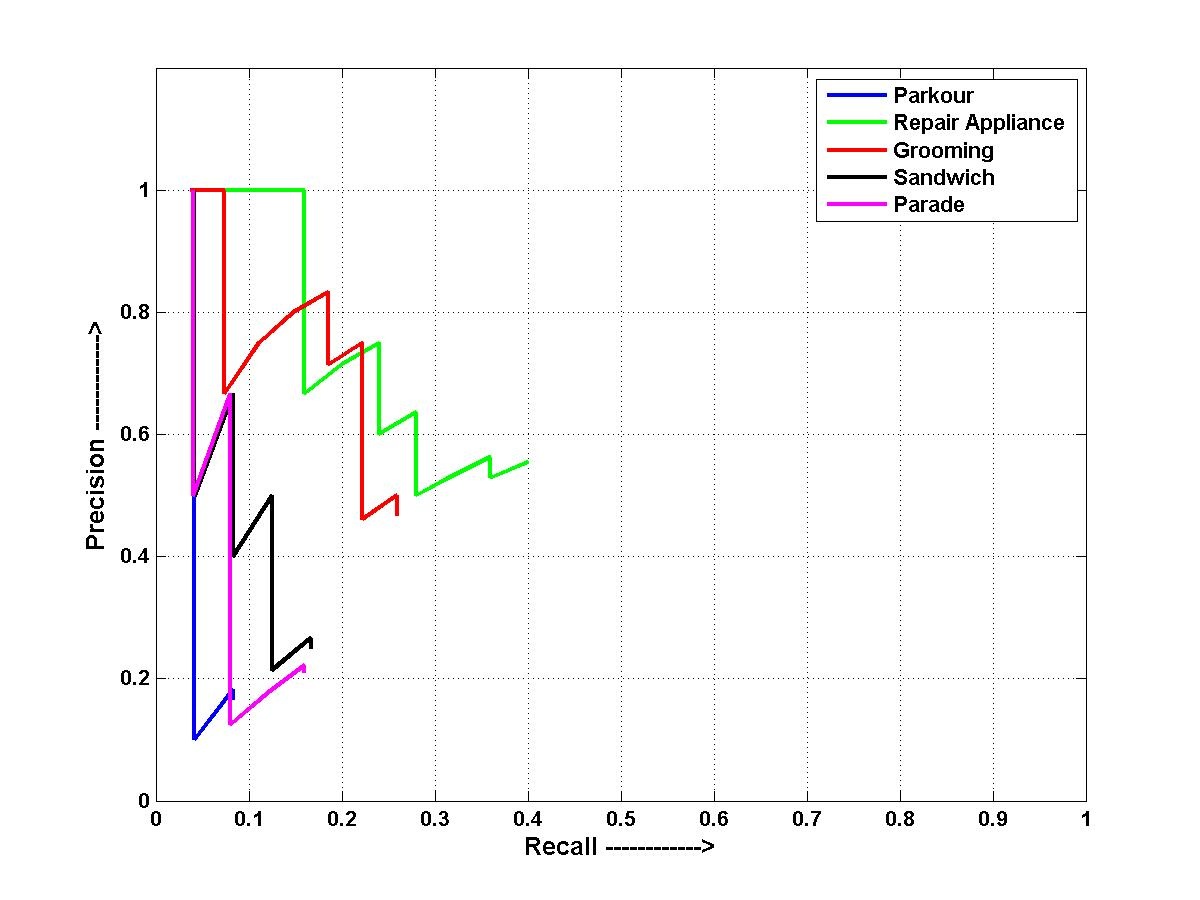}%
\label{dsift}}
\subfloat[SVM-CTS]{\includegraphics[width=2.2in]{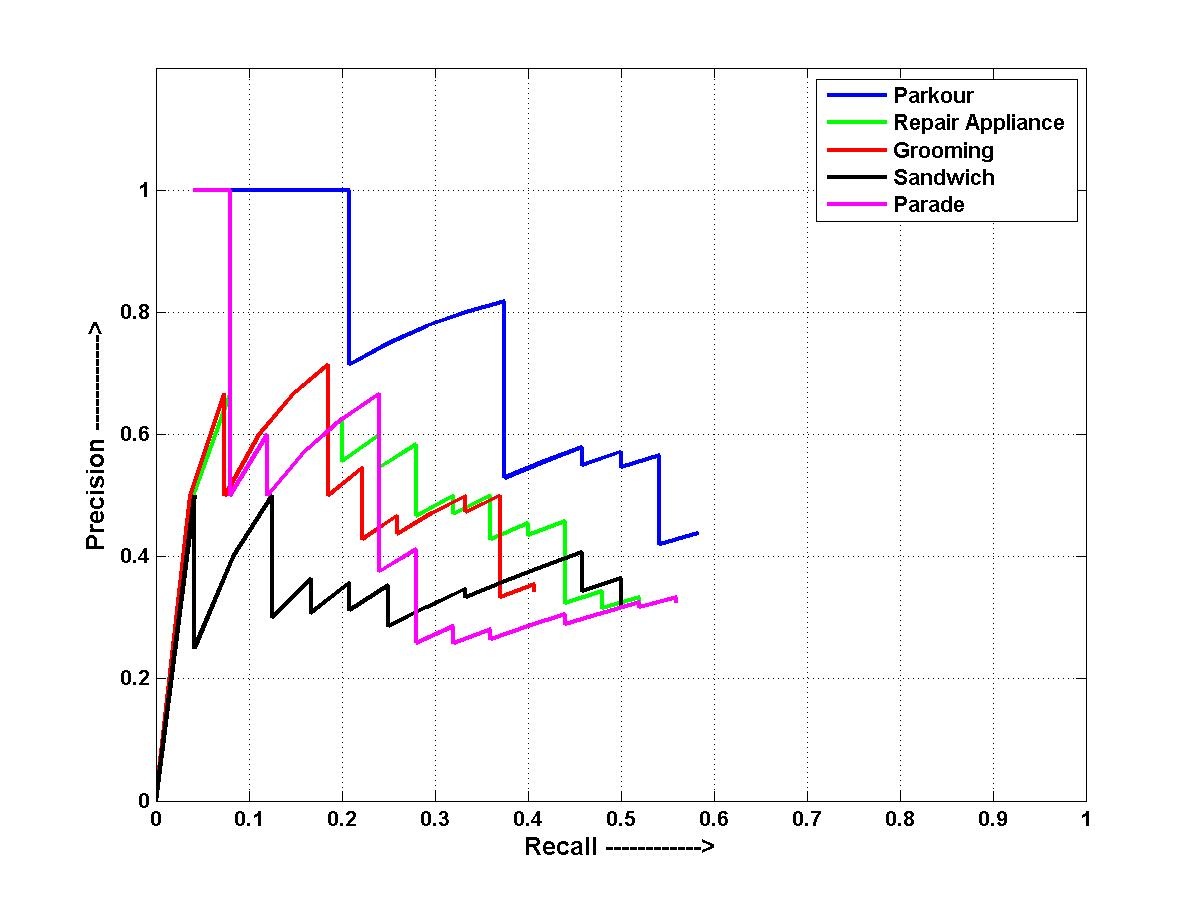}%
\label{cts}}
\caption{Precision-Recall curves}
\label{prc}
\end{figure*}





\item \textit{Video Indexing} :\\
In our experiments, the quality of the obtained annotations are quantified in two different ways:

\begin{enumerate}
\item \textit{Perplexity}, which is given by:

\begin{equation}
Perplexity=\exp\left(\frac{-\sum\limits_{i=1}^{N}\sum\limits_{m=1}^{M_i} ln\ P(d_m|V_i)}{\sum\limits_{i=1}^{N} M_i}\right)
\end{equation} where $N$ represents the number of test videos, $M_i$ the number of words in the caption associated with $V_i$ and $d_m$ represents the $m^{th}$ word of the caption obtained.

Note that higher $P(d_m|V_i)$ implies better annotation as there is a higher probability that the word $d_m$ is associated with the video $V_i$.
From the above formula, we can see that as $P(d_m|V_i)$ increases, the value of \textit{perplexity} decreases which implies that annotation quality and perplexity are \textit{inversely proportional}.

\item \textit{Mean Per-word Precision}, \textit{Mean Per-word Recall} and \textit{F-Score} \cite{hoffman2009easy}:  Per-word recall is defined as the average fraction of videos actually labeled with $w$ that our model annotates with label $w$. Per-word precision is defined as the average fraction of videos that our model annotates with label $w$ that are actually labeled with $w$. F-score is the harmonic mean of precision and recall, and is one metric of overall annotation performance. For our experiments, we considered the top $10$ annotations generated for each video.

We had a total of $9654$ textual words in our dictionary, and computed the per-word metrics for each of these, averaging the results to get the required mean metrics. In some cases, when our model did not annotate any video with the label $w$, we determined the precision for that word by Monte-carlo methods: Generate sets of random annotations for each of the videos and average over the computed precision. If no video in the test set is labeled with $w$, then per-word precision and recall for $w$ are undefined, so we ignored these words in our evaluation.

\end{enumerate}
    
\begin{table}
\vspace{2ex}
\begin{tabular}{c | c | c | c}
\toprule
\textbf{Annotation Length} & \textbf{5 topics} & \textbf{50 topics} & \textbf{100 topics} \\
\midrule
 5 &  33.1512 & 31.2510 & 30.5603\\
 10 & 43.3720 & 41.6222 & 40.5960\\
15 & 53.0023 & 51.0888 & 49.9647\\
20 & 64.2810 & 61.5055 & 60.2332\\
25 & 75.9930 & 72.2725 & 71.1561\\
30 & 87.7312 & 83.1981 & 82.1523\\
35 & 99.4043 & 94.2035 & 93.3511\\
40 & 111.1342 & 105.4465 & 104.6538\\
45 & 122.8281 & 116.7421 & 115.8785\\
50 & 134.4591 & 128.0057 & 127.0357\\
\bottomrule
\end{tabular}
\centering
\caption{Video Indexing: Perplexity for different annotation lengths}
\label{anno}
\end{table}

\begin{table}
\vspace{2ex}
\begin{tabular}{l | c | c | c}
\toprule
\textbf{Model} & \textbf{MPW Precision} & \textbf{MPW Recall} & \textbf{F-Score}\\


\midrule
Corr-LDA-5 & 0.02989 & 0.03066 & 0.03027\\
Corr-LDA-50 & 0.04153 & 0.03871 & 0.04005\\
Corr-LDA-100 & 0.05537 & 0.04844 & 0.05126\\

\bottomrule
\end{tabular}
\centering
\caption{Comparison of Mean Per-word (MPW) Precision, Mean Per-word Recall and F-Score of the Corr-LDA model with different number of topics}
\label{perWord}
\end{table}

It is clear from Table \ref{anno} that the perplexity values reduce with an increase in the number of topics. It is worth recalling that lower values of perplexities imply better annotation quality. Thus, we see that as more and more of the annotations generated are considered, we get higher/poorer values of perplexity implying poorer annotation quality. It is also clear from Table \ref{perWord} that the annotation quality improves with increase in the number of topics. Having obtained perplexity values comparable to \cite{BleiCorr} and from the precision-recall values and F-scores in Table \ref{perWord}, we can see that the Corr-LDA model successfully learns the conditional text-video distributions.
\end{itemize}


\section{Limitations}
We have identified the following limitations which we would like to work upon in our future investigations. Firstly, it is important to note that audio alone cannot distinguish a recording. Several dissimilar videos may contain similar audio words. This problem can be remedied by using visual features of the recordings i.e. video BoW. We expect an increase in the performance/accuracy of the model when visual features are taken into account. Secondly, the textual dictionary is built from the training data and hence, a search-query containing terms outside of it cannot be handled. Various smoothing techniques can be used for this purpose.
\label{sec:limits}


\section{Conclusions and Future work}
\label{sec:concl}
In this paper we have successfully applied the Corr-LDA model to handle the task of content-based video indexing and retrieval. Our experimental results show that the framework learns the correspondences well with the use of just the audio components of the recordings i.e. using the audio bag-of-words alone. Our method is not just restricted to video and text. Rather, it can be applied in a wide variety of scenarios which includes any form of multimedia, for e.g. audio retrieval using images, video retrieval using audio etc., along with corresponding annotations.

Our work provides an arguably promising new direction for research by building and making use of the inherent semantic linkages between text and video thereby obviating the need for frame-by-frame search. In the future, we would like to further our method by using the video bag-of-words for the stated task. Also, a combination of tag-based query-search and content-based semantic linkages can be used to gain a better performance, by incorporating the merits of both the methods.

\bibliography{refs}
\bibliographystyle{aaai}
\end{document}